\documentclass[aps,preprint,showpacs,nofootinbib,12pt]{revtex4}
\usepackage{graphicx}
\usepackage{epsfig}
\usepackage[dvips]{color}

\begin{document}

\title{Is $Z_b(10610)$ a Molecular State?}

\author{ Hong-Wei Ke$^{1}$   \footnote{khw020056@hotmail.com},
         Xue-Qian Li $^{2}$  \footnote{lixq@nankai.edu.cn},
          Yan-Liang Shi$^{2}$   \footnote{shiyl@mail.nankai.edu.cn},
Guo-Li Wang$^3$  \footnote{gl\_wang@hit.edu.cn}
         and
          Xu-Hao Yuan$^4$
         \footnote{yuanxh@tsinghua.edu.cn}
       }

\affiliation{
  $^{1}$ School of Science, Tianjin University, Tianjin 300072, China \\
  $^{2}$ School of Physics, Nankai University, Tianjin 300071, China\\
  $^{3}$ Department of Physics, Harbin Institute of
Technology, Harbin, 150001, China\\
 $^4$ Center for High Energy Physics,
Department of Engineering Physics, Tsinghua University, Beijing
100084, China }

\begin{abstract}
\noindent Whether molecular states indeed exist in nature has been
disputed for a long time. Several new resonances have been observed
in the recent experiments and they seem to be of exotic structures
and some of them have been proposed to be molecular states. The very
recent observation of $Z_b(10610)[(10608.4\pm 2.0)$ MeV] and
$Z_b(10650)[(10653.2\pm 1.5)$ MeV] encourages the interpretation of
multi-quark states. In the  Beter-Salpeter (BS) approach, we study
the possibility if two heavy mesons can form a molecular state  by
exchanging light mesons. Our results indicate that two heavy mesons
can form an isospin singlet ($I=0$) bound state but cannot form an
isospin triplet ($I=1$) when the contribution of $\sigma-$ exchange
is reasonably small, i.e. as the coupling  of $\sigma$ with mesons
$g_{\sigma}$ takes the value given in previous literatures. Thus we
conclude that the newly observed $Z_b(10610)$ should not be a
molecular state, but a tetraquark state instead, at most, the
fraction of the molecular state in the physical resonance
$Z_b(10610)$ is tiny.

\end{abstract}


\maketitle

\section{introduction}
The naive non-relativistic quark model\cite{GellMann:1964nj} tells
us that a meson contains a quark and an anti-quark while a baryon
is composed of three valence quarks. However, it is also noted
that the exotic hybrid, glueball and multi-quark states are not
excluded by the $SU(3)$ quark model. Until now no any meson or
baryon has been confirmed as an exotic state even though several
resonances are considered to be exotic, namely have different
component-structure from regular hadrons. There is not any
principle to forbid their existence, and in fact the discussion
about the exotic states has never stopped. For example the exotic
structure of $f_0(980)$ and $a_0(980)$ is still under intense
dispute \cite{f01,f02,f03,f04,f011}. Since the conventional $q\bar
q$ structure cannot well fit experimental data of certain
modes\cite{:2009cm,Ke:2009ed}, some authors proposed that they may
be molecular states made of two color-singlet mesons or
tetraquarks composed of a color-anti-triplet diquark and a
color-triplet anti-diquark \cite{f01,f02,f03}, .

Recently, a series of charmonium-like resonances have successively
been experimentally observed, such as $X(3872)$\cite{Choi:2003ue},
$X(3940)$\cite{Abe:2007jn}, $Y(3940)$\cite{Choi:2005},
$Z(4430)^{\pm}$\cite{Choi:2007wga}  and several bottomoniun-like
states are also discovered by the Belle and Babar collaborations,
such as $Z_b(10610)$ and $Z_b(10650)$ \cite{Collaboration:2011gj}.
It is noted that there is almost no room in the regular
representations of $O(3)\otimes SU_f(3)\otimes SU_s(2)$ to
accommodate those newly observed resonances. Especially these
particles $Z(4430)$, $Z_b(10610)$ and $Z_b(10650)$ with non-zero
charge cannot be understood in the conventional $Q\bar Q$
structure if they are real resonances
\cite{Zhang:2011jja,Sun:2011uh,Bondar:2011ev,Cleven:2011gp,Navarra:2011xa}.
Concretely the mass of $Z_b$ is between the mass of $\Upsilon(4S)$
and $\Upsilon(5S)$, so it should have one b (quark) and one $\bar
b$ (anti-quark) but it by no means is a bottomonium because of its
electric charge. Apparently if it is verified to be a resonance it
should be a multi-quark state, namely a molecular state or
tetraquark should be the preferred choice since they are the
simplest extension beyond the regular $Q\bar Q$ structure.

Guo et al. explored possible $K\bar K$ bound states  in the
Bethe-Salpeter framework and found that the bound state could exist
\cite{Guo:2007mm}. In the same framework the resonance
$B^*_{s0}(5725)$ was considered as a $B\bar K$ molecular
state\cite{Feng:2011zzb}. Whether two heavy mesons can form a
molecular state has not been thoroughly investigated in those
works\cite{Feng:2011zzb,Guo:2007mm}. Thus in this work we will try
to study the question in the Bethe-Salpeter (BS) approach where the
relativistic corrections are automatically included. Besides the
bound states of two pseudoscalars we also  explore the bound states
which contain one or two vectors. For the aim we need to deduce
corresponding BS equations. Since  in this work we only concern the
ground states,  the orbital angular momentum between two constituent
mesons is zero ($l=0)$. For a molecular state whose constituents are
one pseudoscalar and one vector,  its $J^{PC}$ is $1^+$. For the
molecular states which consist of two vector mesons their $J^{PC}$
may be $0^+$, $1^+$ and $2^+$. We only deduce the BS equation for
the bound state with $J^P=0^+$  because it should be the lightest
molecular state and more favorable in the nature. As a matter of
fact, the Lorentz structures of the others are very complicated and
will be explored in our later works.

When we solve the BS equations for these molecular states, two
approximations are needed: the   ladder approximation and
instantaneous approximation. In general the energy exchange between
the constituents is small compared to $\Lambda_{QCD}$ which
characterizes the binding energy scale of the constituent mesons, so
that the instantaneous approximation is applicable. The simple
ladder approximation has been employed all along in the history, but
recently some works suggest that the cross-ladder should be
included\cite{Gross:1982nz,Nieuwenhuis:1996mc}. Since our goal is to
study the possibility of formation of molecular states instead of
making precise theoretical predictions on the spectra  we will still
use the simple ladder approximation to gain our qualitative
conclusion.

After this introduction we derive the BS equations for the $0^+$ and
$1^+$ molecular states. Then in section III we present our numerical
results of the binding energies along with explicitly displaying all
input parameters. Section IV is devoted to a brief summary.

\section{The Bethe-Salpeter formalism}
In this section we will deduce the BS equations for the $0^+$ and
$1^+$ molecular bound states.

\subsection{The bound state ($0^+$) composed of two pseudoscalar mesons}
In Ref. \cite{Guo:2007mm,Feng:2011zzb} the BS equation of a bound
state of two pseudoscalars was deduced. Since in this work we need
to use the corresponding formulas for analyzing the spectra of the
molecular states containing two pseudoscalars, let us briefly review
the main contents of Ref. \cite{Guo:2007mm,Feng:2011zzb}, then in
the following subsections, we will step forward to deduce the BS
equations for one-vector-one-pseudoscalar bound states and
two-vector bound states .

The BS wave function for the bound state $|\mathcal{P}\rangle$ of
two pseudoscalar  mesons can be defined  as following:
\begin{eqnarray}\label{definition-BS1} {\chi}_{{}_\mathcal{P}}^{}(x_1,x_2) = \langle 0 | {\rm
T}\,\phi_1(x_1)\phi_2(x_2) | \mathcal{P} \rangle = e^{-iPX}
{\chi}_{{}_\mathcal{P}}^{}(x)\,,
\end{eqnarray}
where $\phi_1(x_1)$ and $\phi_2(x_2)$ are the field operators of two
mesons, respectively, $P$ denotes the total momentum of the bound
state, the relative coordinate $x$ and the center of mass coordinate
$X$ are
\begin{eqnarray} X=\eta_1 x_1 + \eta_2 x_2\,,\quad x = x_1 -
x_2 \,, \end{eqnarray} where $\eta_i = m_i/(m_1+m_2)$ and $m_i\,
(i=1,2)$ is the mass of the $i$-th constituent meson. The equation
for the BS wave function can be derived from a four-point Green
function,
\begin{eqnarray}\label{four-point-green-function1}
S(x_1,x_2;y_2,y_1) = \langle 0 | {\rm T}\,\phi_{1}(x_1)\phi_2(x_2)
(\phi_1(y_1)\phi_2(y_2))^\dag | 0 \rangle \,.
 \end{eqnarray}
To obtain the BS equation, we express the above four-point Green
function in terms of the four-point truncated irreducible kernel
$\overline K$\,,
\begin{eqnarray}
&&S(x_1,x_2;y_2,y_1) = S_{(0)}(x_1,x_2;y_2,y_1) \nonumber\\
&& \quad + \int d^4u_1 d^4 u_2 d^4 v_1 d^4 v_2\,
S_{(0)}(x_1,x_2;u_2,u_1) {\overline K} (u_1,u_2;v_2,v_1)
S(v_1,v_2;y_2,y_1) \,, \label{irreducible-rep1}
\end{eqnarray}
where $S_{(0)}$ is related to the forward
scattering disconnected four-point amplitude,
\begin{eqnarray}
S_{(0)}(x_1,x_2;y_2,y_1) = \Delta_1(x_1,y_1)\Delta_{2}(x_2,y_2)
\,,
\end{eqnarray}
and $\Delta_i(x_i, y_i)$ is the complete propagator of the $i$-th
meson. Here we have
\begin{eqnarray}\label{bs-equation-momentum1}
 \Delta_1^{-1}(p_1,m_1)\Delta_{2}^{-1}(p_2,m_2){\chi}_{{}_\mathcal{P}}^{}(p) = \int {d^4p'\over
(2\pi)^4} {\overline K_1}({ p},{
p}'){\chi}_{{}_\mathcal{P}}^{}(p')\,,
 \end{eqnarray}
where  $ \Delta_1=\frac{i}{p_1^2-m_1^2}$, $
\Delta_2=\frac{i}{p_2^2-m_2^2}$ and $ \overline{K_1}(p,p') = -\,i
 ~c_I^{} \, g^2
{(p_1+p_1')\cdot(p_2+p_2')+(p_1^2-p_1'{}^2)(p_2^2-p_2'{}^2)/M_{\rm
V}^2 \over (p_1-p_1')^2-M_{\rm V}^2} $ which has been deduced in
Ref. \cite{Guo:2007mm,Feng:2011zzb}.

The relative momenta and the total momentum of the bound state in
the equations are defined as
\begin{eqnarray} p = \eta_2p_1 -
\eta_1p_2\,,\quad p' = \eta_2p'_1 - \eta_1p'_2\,,\quad P = p_1 +
p_2 = p'_1 + p'_2 \,. \label{momentum-transform1}
\end{eqnarray}

Directly solving the BS equation (\ref{bs-equation-momentum1}) is
extremely difficult. In general one needs to use the so-called
instantaneous approximation:
$\overline{K_1}(p,p')={K_1}(\mathbf{p},\mathbf{p}')$ by which the BS
equation can be reduced to
\begin{eqnarray} \label{3-dim-BS1}
{E^2-(E_1+E_2)^2\over (E_1+E_2)/E_1E_2}
\widetilde\chi_{{}_\mathcal{P}}^{}({\bf p}) ={i\over
2}\int{d^3\mathbf{p}'\over(2\pi)^3}\, {\overline{} K_{1V(S)}}({\bf
p},{\bf p}')\widetilde\chi_{{}_\mathcal{P}}^{}({\bf p}')F(\bf
p-\bf p')^2 \,,
\end{eqnarray}
where $E_i \equiv \sqrt{{\bf p}^2 + m_i^2}$, $E=P^0$, and the
equal-time wave function is defined as $ \widetilde\chi_{_P}({\bf
p})= \int dp^0 \, \chi_{_P}(p) \,. $ Since the constituent meson is
not a point particle  a form factor at each interaction vertex among
hadrons must be introduced to reflect the finite-size effects of
these hadrons. The form factor is assumed to be in the following
form:
\begin{eqnarray} \label{form-factor} F({\bf k}) = {2\Lambda^2 -
M_{\rm V}^2 \over 2\Lambda^2 + {\bf k}^2}\,,\quad {\bf k} = {\bf
p}-{\bf p}' \,,
\end{eqnarray} where $\Lambda$ is a cutoff parameter.
For  exchange of a light vector between the mesons, the kernel is
\begin{eqnarray} K_{1V}({\bf p},{\bf
p}') = i c_I\,{g_{_{PVP}}g_{_{PVP}}'}\, {({\bf p}+{\bf p}')^2 +
4\eta_1\eta_2 E^2 + ({\bf p}^2-{\bf p}'{}^2)^2/M_{\rm V}^2 \over
({\bf p}-{\bf p}')^2 + M_{\rm V}^2} \,.
\label{potential-with-isospin-factor}
\end{eqnarray}

The exchanged mesons between two pseudoscalars are vector mesons,
obviously we only need to keep the lightest vector mesons $\rho$ and
$\omega$ for taking the dominant contributions into account
\cite{Guo:2007mm,Feng:2011zzb}. When the bound state is an
isospin-scalar $c_I=3$ for $\rho\,(\pi)$ and $c_I=1$ for
$\omega\,(\sigma)$. When the bound state is an isospin-vector
$c_I=-1$ for $\rho\,(\pi)$ and $c_I=1$ for $\omega\,(\sigma)$.

\subsection{The bound state($1^+$) composed of a pseudoscalar and a vector}
We can define the BS wave function for the bound state
$|\mathcal{V}\rangle$ composed of one pseudoscalar and a vector
mesons as following:
\begin{eqnarray}\label{definition-BS2} {\chi^\mu}_{{}_\mathcal{V}}^{}(x_1,x_2) = \langle 0 | {\rm
T}\,\phi_1(x_1)\phi^\mu_2(x_2) | \mathcal{V} \rangle = e^{-iPX}
{\chi^\mu}_{{}_\mathcal{V}}^{}(x)\,,
\end{eqnarray}
where $\phi_1(x_1)$ and $\phi^\mu_2(x_2)$ are respectively the
field operators of the two  mesons. The equation for the BS wave
function should be derived from a four-point Green function,
\begin{eqnarray}\label{four-point-green-function2}
S^{\mu\nu}(x_1,x_2;y_2,y_1) = \langle 0 | {\rm
T}\,\phi^\mu_{1}(x_1)\phi_2(x_2) (\phi^\nu_1(y_1)\phi_2(y_2))^\dag
| 0 \rangle \,.
 \end{eqnarray}
To obtain the corresponding BS equation, we write the above
four-point Green function in terms of the four-point truncated
irreducible kernel $\overline K$\,,
\begin{eqnarray}
&&S^{\mu\nu}(x_1,x_2;y_2,y_1) = S^{\mu\nu}_{(0)}(x_1,x_2;y_2,y_1) \nonumber\\
&& \quad + \int d^4u_1 d^4 u_2 d^4 v_1 d^4 v_2\,
S^{\mu\alpha}_{(0)}(x_1,x_2;u_2,u_1) {\overline K_{\alpha\beta}}
(u_1,u_2;v_2,v_1) S^{\beta\nu}(v_1,v_2;y_2,y_1) \,,
\label{irreducible-rep3}
\end{eqnarray}
where $S_{(0)}$ is related to the forward-scattering disconnected
four-point amplitude,
\begin{eqnarray} S^{\mu\alpha}_{(0)}(x_1,x_2;y_2,y_1) =
\Delta_1(x_1,y_1)\Delta^{\mu\alpha}_2(x_2,y_2) \,,
\end{eqnarray}
and $\Delta_i(x_i, y_i)$ is the full propagator of the $i$-th
particle,
\begin{eqnarray}\label{bs-equation-momentum2}
 \Delta_1^{-1}(p_1,m_1)\Delta_{2\mu\alpha}^{-1}(p_2,m_2){\chi^\mu}_{{}_\mathcal{V}}^{}(p) = \int {d^4p'\over
(2\pi)^4} {\overline K_{\alpha\beta}}({ p},{
p}'){\chi^\beta}_{{}_\mathcal{V}}^{}(p')\,.
\end{eqnarray}
Here $ \Delta_1=\frac{i}{p_1^2-m_1^2}$ and $
\Delta_{2\mu\alpha}=\frac{i}{p_2^2-m_2^2}(\frac{p_{2\mu}p_{2\alpha}}{m_2^2}-g_{\mu\alpha})$
are the propagators of pseudoscalar and vector mesons.

From Eq. (\ref{bs-equation-momentum2}) one can obtain
\begin{eqnarray}\label{bs-equation-momentum2p}
\frac{-1}{(p_1^2-m_1^2)(p_2^2-m_2^2)}{\chi^\mu}_{{}_\mathcal{V}}^{}(p)
= \int {d^4p'\over (2\pi)^4} {\overline K_{\alpha\beta}}({ p},{
p}'){\chi^\beta}_{{}_\mathcal{V}}^{}(p')(\frac{p_{2}^\mu
p_{2}^\alpha}{m_2^2}-g^{\mu\alpha})\,.
 \end{eqnarray}

We write
${\chi^\mu}_{{}_\mathcal{V}}^{}(p)={\chi}_{{}_\mathcal{V}}^{}(p)\epsilon^\mu$,
then multiply an $\epsilon^*_\mu$ on both sides. Summing over the
polarizations we deduce a new equation
\begin{eqnarray}\label{bs-equation-momentum2pp}
\frac{-1}{(p_1^2-m_1^2)(p_2^2-m_2^2)}{\chi}_{{}_\mathcal{V}}^{}(p)
= \int {d^4p'\over (2\pi)^4} {\overline K_{\alpha\beta}}({ p},{
p}'){\chi}_{{}_\mathcal{V}}^{}(p')(\frac{p_{2}^\mu
p_{2}^\alpha}{m_2^2}-g^{\mu\alpha})(\frac{p'_\mu
p'^\beta}{p'^2}-g_\mu^\beta)\,.
 \end{eqnarray}
With the Feynman diagrams depicted in Fig.1 and the effective
interactions\cite{Guo:2007mm,Zhang:2011jja} we eventually obtain
$$\overline K_{\alpha\beta}=g_{_{PVP}}g_{_{VVV}}\frac{\frac{q^\lambda q^\sigma}{q^2}-g^{\lambda\sigma}}{q^2-M_V^2}(p_1+p_1')_\lambda[-g_{\alpha\beta}
(p_2+p_2')_\sigma+{p_2'}_\alpha g_{\sigma\beta}+{p_2}_\beta
g_{\sigma\alpha}].
$$

Defining $ \overline K_2(p,p')={\overline K_{\alpha\beta}}({ p},{
p}')(\frac{p_{2}^\mu
p_{2}^\alpha}{m_2^2}-g^{\mu\alpha})(\frac{p'_\mu
p'^\beta}{p'^2}-g_\mu^\beta)$ and employing the instantaneous
approximation we obtain the BS equation similar to Eq.
(\ref{bs-equation-momentum1}) and (\ref{3-dim-BS1}) but with a
different kernel,
\begin{eqnarray} \label{3-dim-BS2}
{E^2-(E_1+E_2)^2\over (E_1+E_2)/E_1E_2}
\widetilde\chi_{{}_\mathcal{V}}^{}({\bf p}) ={i\over
2}\int{d^3\mathbf{p}'\over(2\pi)^3}\, {\overline{} K_{2V(S)}}({\bf
p},{\bf p}')\widetilde\chi_{{}_\mathcal{V}}^{}({\bf p}')F(\bf
p-\bf p')^2 \,.
\end{eqnarray}

\begin{center}
\begin{figure}[htb]
\begin{tabular}{cc}
\scalebox{0.5}{\includegraphics{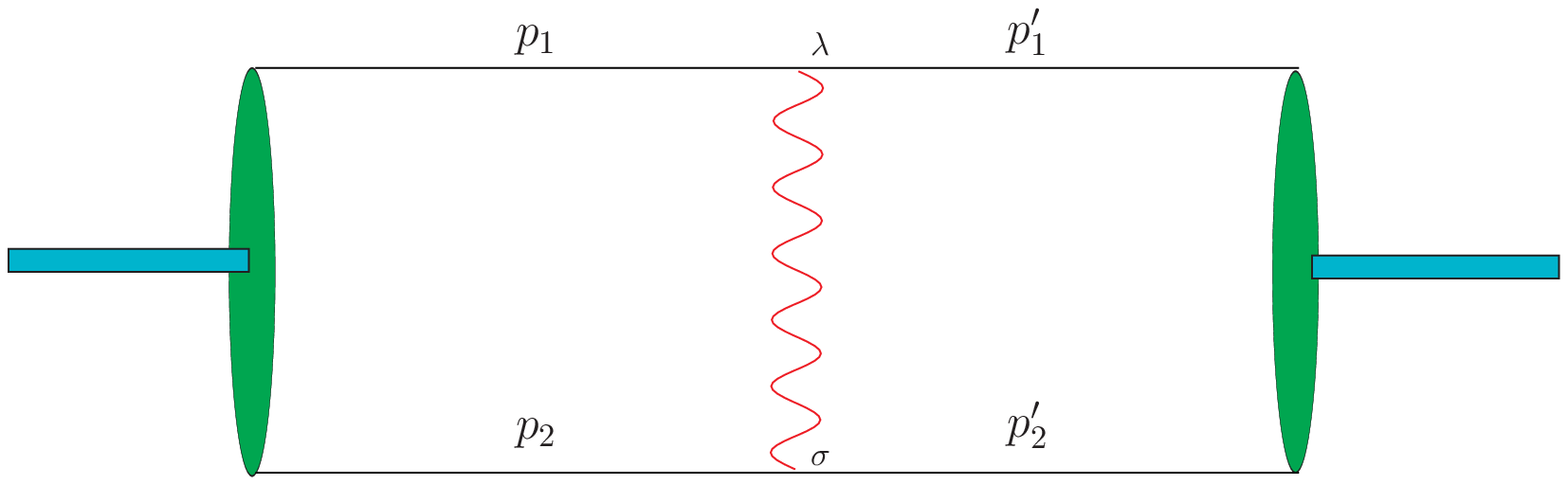}}\scalebox{0.5}{\includegraphics{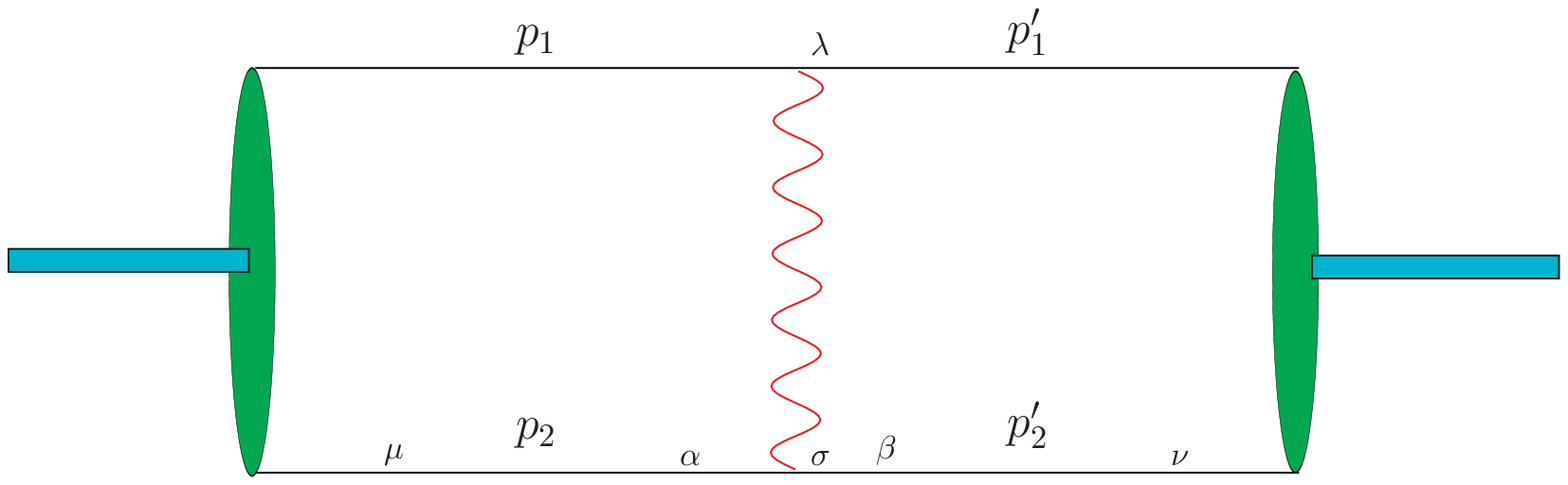}}\\
(a)\,\,\,\,\,\,\,\,\,\,\,\,\,\,\,\,\,\,\,\,\,\,\,\,\,\,\,\,\,\,\,\,\,\,\,\,
\,\,\,\,\,\,\,\,\,\,\,\,\,\,\,\,\,\,\,\,\,\,\,\,\,\,\,\,\,\,\,\,\,\,\,\,
\,\,\,\,\,\,\,\,\,\,\,\,\,\,\,\,\,\,\,\,\,\,\,\,\,\,\,\,\,\,\,\,\,\,\,\,
\,\,\,\,\,\,\,\,\,\,\,\,\,\,\, (b)
\end{tabular}
\caption{(a) A bound state composed of two pseudoscalars (b) a
bound state composed of a pseudoscalar and a vector. In the two
Feynman diagrams $\rho$ and $\omega$ are exchanged.}\label{DM2}
\end{figure}
\end{center}

\begin{center}
\begin{figure}[htb]
\begin{tabular}{cc}
\scalebox{0.5}{\includegraphics{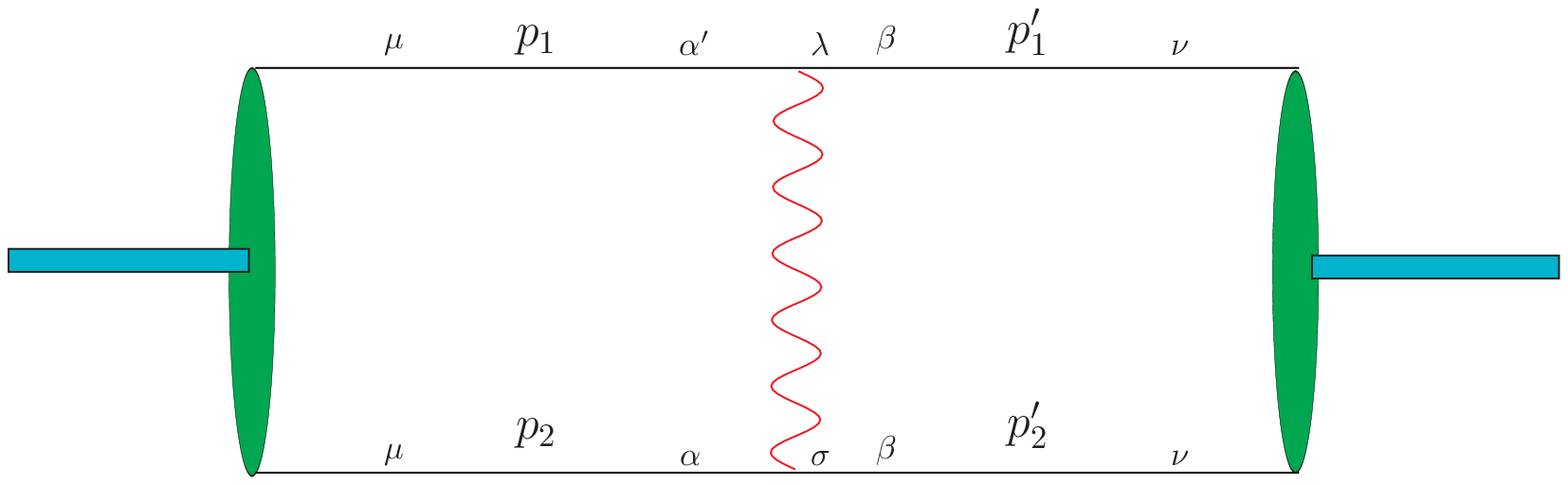}}\scalebox{0.5}{\includegraphics{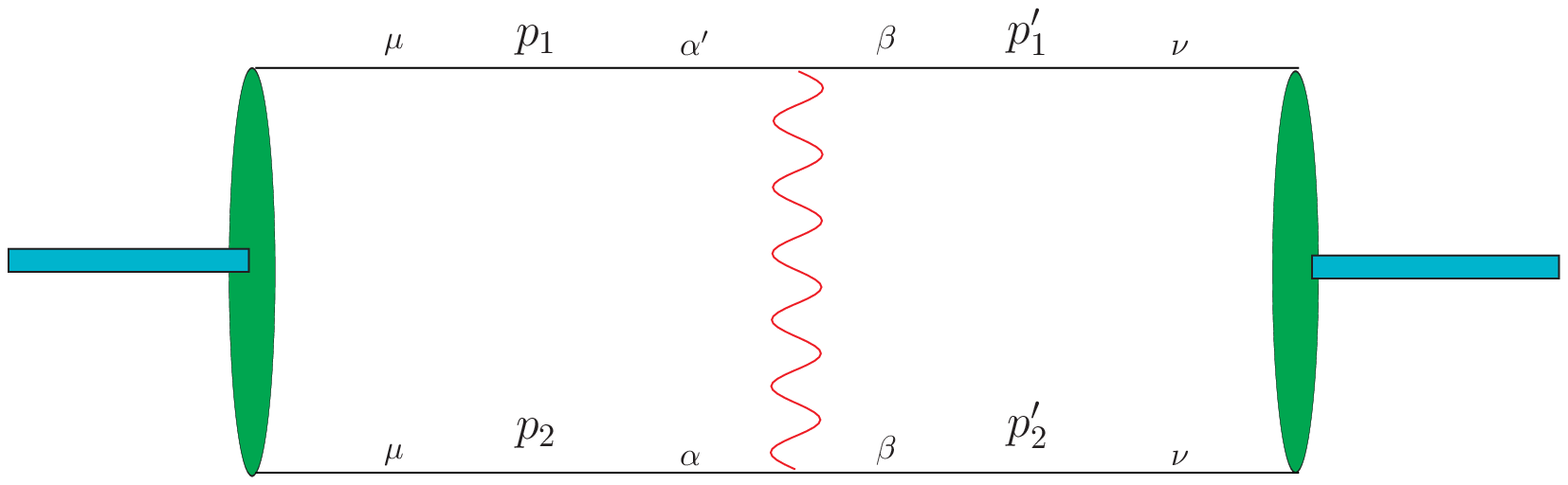}}
\\
(a)\,\,\,\,\,\,\,\,\,\,\,\,\,\,\,\,\,\,\,\,\,\,\,\,\,\,\,\,\,\,\,\,\,\,\,
\,\,\,\,\,\,\,\,\,\,\,\,\,\,\,\,\,\,\,\,\,\,\,\,\,\,\,\,\,\,\,\,\,\,\,\,
\,\,\,\,\,\,\,\,\,\,\,\,\,\,\,\,\,\,\,\,\,\,\,\,\,\,\,\,\,\,\,\,\,\,\,\,
\,\,\,\,\,\,\,\,\,\,\,\,\,\,\, (b)
\end{tabular}
\caption{A bound state composed of two vectors. (a) $\rho$ and
$\omega$ are exchanged  (b) $\pi$ is exchanged}\label{DM2}
\end{figure}
\end{center}

\subsection{The bound state($0^+$) composed of two vector mesons}

The quantum number $J^P$ of the bound state composed of two vectors
can be $0^+$, $1^+$ and $2^+$. As aforementioned, since $0^+$ is
more favorable in nature and its Lorentz structure is relatively
simpler than the two others, in this work we only study the $0^+$
bound states and define the corresponding BS wave function
$|\mathcal{S}\rangle$ as following:
\begin{eqnarray}\label{definition-BS} {\chi}_{{}_\mathcal{S}}^{}(x_1,x_2) = \langle 0 | {\rm
T}\,\phi_{1\mu}(x_1)\phi^\mu_2(x_2) | \mathcal{S} \rangle =
e^{-iPX} {\chi}_{{}_\mathcal{S}}^{}(x)\,.
\end{eqnarray}
The equation for the BS wave function
can be derived from a four-point Green function,
\begin{eqnarray}\label{four-point-green-function}
S(x_1,x_2;y_2,y_1) = \langle 0 | {\rm
T}\,\phi_{1\mu}(x_1)\phi^\mu_2(x_2)
(\phi_{1\nu}(y_1)\phi^\nu_2(y_2))^\dag | 0 \rangle \,.
 \end{eqnarray}
In analog to the procedures used in last subsection,  we obtain
\begin{eqnarray}\label{bs-equation-momentum}
 \Delta_{1\mu\alpha}^{-1}(p_1,m_1)\Delta_{2}^{-1\mu\alpha'}(p_2,m_2){\chi}_{{}_\mathcal{S}}^{}(p) = \int {d^4p'\over
(2\pi)^4} {\overline K_{\alpha}^{\,\,\alpha'}}({ p},{
p}'){\chi}_{{}_\mathcal{S}}^{}(p')\,,
 \end{eqnarray}
where
$\Delta_{1\mu\alpha}=\frac{i}{p_1^2-m_1^2}(\frac{p_{1\mu}p_{1\alpha}}{m_2^2}-g_{\mu\alpha})$
and
$\Delta_{2\mu\alpha'}=\frac{i}{p_2^2-m_2^2}(\frac{p_{2\mu}p_{2\alpha'}}{m_2^2}-g_{\mu\alpha'})$.

From Eq. (\ref{bs-equation-momentum2}) one can obtain
\begin{eqnarray}\label{bs-equation-momentum2p}
\frac{-1}{(p_1^2-m_1^2)(p_2^2-m_2^2)}{\chi^\mu}_{{}_\mathcal{S}}^{}(p)
= \int {d^4p'\over (2\pi)^4} {\overline
K_{\alpha}^{\,\,\alpha'}}({ p},{
p}'){\chi_\mathcal{S}}_{{}}^{}(p')(\frac{p_{2}^\mu
p_{2}^\alpha}{m_2^2}-g^{\mu\alpha})(\frac{{p_{1}}_\mu
{p_{1}}_{\alpha'}}{m_2^2}-g_{\mu\alpha'}).
 \end{eqnarray}
With the Feynman diagrams depicted in Fig.2 and the effective
interaction\cite{Feng:2011zzb,Guo:2007mm,Zhang:2006ix,Ke:2010aw} we
can obtain
$$\overline K_{V\alpha}^{\alpha'}=g_{_{VVV}}g_{_{VVV}}'\frac{\frac{q^\lambda q^\sigma}{q^2}-g^{\lambda\sigma}}{q^2-M_V^2}[-g^{\alpha'\beta}
(p_1+p_1')_\lambda+{p_1'}^{\alpha'}
g_{\lambda}^{\beta}+{p_1}^\beta
g_{\lambda}^{\alpha'}][-g_{\alpha\beta}
(p_2+p_2')_\sigma+{p_2'}_\alpha g_{\sigma\beta}+{p_2}_\beta
g_{\sigma\alpha}]
$$
and $$\overline
K_{P\alpha}^{\alpha'}=g_{_{VPV}}g_{_{VPV}}'\varepsilon^{\alpha'\beta\sigma\rho}q_\sigma
p_{1\rho}\varepsilon_{\alpha\beta\sigma'\rho'}q^{\sigma'}
p_2^{\rho'}.
$$

Defining $ \overline K_{3V(P,S)}(p,p')={\overline
K_{V(P,S)\alpha}^{\,\,\alpha'}}({ p},{ p}')(\frac{p_{2}^\mu
p_{2}^\alpha}{m_2^2}-g^{\mu\alpha})(\frac{{p_{1}}_\mu
{p_{1}}_{\alpha'}}{m_2^2}-g_{\mu\alpha'})$ we derive the BS
equation which is similar to Eqs. (\ref{bs-equation-momentum1})
and (\ref{3-dim-BS1}) but possesses a different kernel,
\begin{eqnarray} \label{3-dim-BS3}
{E^2-(E_1+E_2)^2\over (E_1+E_2)/E_1E_2}
\widetilde\chi_{{}_\mathcal{S}}^{}({\bf p}) ={i\over
2}\int{d^3\mathbf{p}'\over(2\pi)^3}\, {\overline{}
K_{3V(P,S)}}({\bf p},{\bf
p}')\widetilde\chi_{{}_\mathcal{S}}^{}({\bf p}')F(\bf p-\bf p')^2
\,.
\end{eqnarray}

\section{Numerical results}
Solving the BS equations  Eq. (\ref{3-dim-BS1}), (\ref{3-dim-BS2})
and (\ref{3-dim-BS3}), we obtain the eigenvalues for each state. To
determine whether such a state composed of two heavy mesons is a
bound state, a criterion must be set. That is: if the eigenvalue of
the bound state obtained by solving the BS equation is negative,
namely the total mass of the system is lower than the sum of the
masses of the two constituents, such a system is considered as a
bound state, i.e. a hadronic molecule which may exist in the nature.

Now let us solve the BS equations. Since the function
$\widetilde\chi_\mathcal{P}(\mathbf{p})$,
$\widetilde\chi_\mathcal{V}(\mathbf{p})$,
$\widetilde\chi_\mathcal{S}(\mathbf{p})$ only depends on the norm
of the three-momentum we may first integrate over the azimuthal
angle of the function in (\ref{3-dim-BS1}), (\ref{3-dim-BS2}) or
(\ref{3-dim-BS3})
$$\frac{i}{2}\int{d^3\mathbf{p}'\over(2\pi)^3}\, {\overline{} K}({\bf p},{\bf
p}')F(\bf p-\bf p')^2,  $$ to obtain a new form $U(|\mathbf{p}|,|\mathbf{p}'|)$,
then the BS equation turns into a
one-dimension integral equation
\begin{eqnarray} \label{3-dim-BS4}
\widetilde\chi({\bf |p|}) -{(E_1+E_2)/E_1E_2\over E^2-(E_1+E_2)^2
}\int{d \mathbf{|p}'|}\, {\overline{} U}({\bf |p|},{\bf
|p}'|)\widetilde\chi({\bf |p}'|) =0.
\end{eqnarray}

Letting $\bf |p|(|p'|)$  take ordered $n$ discrete values and the
gap between two adjacent values be $\Delta \bf p$, then $\chi({\bf
|p|})$ can be arranged as a column matrix and the coefficients
constitute an $n\times n$ matrix $M$. The scalar equation of $M$
would determine the $n$ eigenvalues $E_n$ and the smallest one
should correspond to the binding energy of the ground state. In
our calculation we take $n=151$ and the first discrete value of
$|\mathbf{p}|=0.001$ GeV and the largest $|\mathbf{p}|$ is $2$
GeV.

In our calculation we need to determine the values of the
parameters $\Lambda$, $g_{_{PVP}},g_{_{VVV}},g_{_{PSP}}$ and
$g_{_{VSV}}$. By fitting the data of the decay processes where the
couplings $KK\rho$ and$BB\rho$ are concerned, the authors of
Ref.\cite{Feng:2011zzb} obtained the cutoff parameter $\Lambda$
which can vary from 1GeV to 4GeV. Their strategy was to fit the
measured decay widths in terms of the theoretically derived
formulas with $\Lambda$, thus they obtained the rather wide range
for $\Lambda$. Indeed such a wide range reduces our theoretical
prediction power. Instead, in Ref. \cite{Cheng:2004ru} the authors
suggested a relation: $\Lambda=m+\alpha \Lambda_{QCD}$ where $m$
is the mass of the exchanged meson, $\alpha$ is a number of $O(1)$
and $\Lambda_{QCD}=220$ MeV i.e. $\Lambda\sim 1$GeV for exchanging
$\rho$ or $\omega$. Since there are uncertainties on $\Lambda$ we
will choose $\Lambda=1.5$ GeV, $\Lambda=2.5$ GeV  and
$\Lambda=3.5$ GeV to calculate the eigenvalues and compare the
results. The coupling constants $g_{BB\rho}=g_{KK\rho}=3$ are
fixed by fitting data \cite{Zhang:2006ix}. Considering the flavor
$SU(3)$ symmetry and heavy quark spin symmetry we believe that
$g_{BB\rho(\omega)}=g_{DD\rho(\omega)}=g_{BB^*\rho(\omega)}=g_{DD^*\rho(\omega)}=g_{B^*B^*\rho(\omega)}=g_{D^*D^*\rho(\omega)}=g$
should be a not-bad approximation. Instead of setting $g=3$, we
also let $g$ take various values from 2 to 4 to investigate
dependence of the eigenvalues on those coupling constants.

The masses of the concerned constituent mesons are directly taken
from the databook \cite{PDG10} as: $m_D=1.865$ GeV, $m_{D^*}=2.007$
GeV, $m_B=5.279$ GeV and $m_{B^*}=5.325$ GeV .
\subsection{The exchanged mesons are $\rho$ and $\omega$}



It is noted that in those tables there are many  places symbolized
by $``-"$ which means such bound states cannot exist because either
the BS equation with the corresponding parameters has no solution at
all, or the derived eigen-energy is not negative, namely the mass of
the supposed-to-be bound state is larger than the sum of the two
constituent mesons.

We calculate the eigenvalues of these  bound states of $D\bar
D(0^+)$, $D\bar D^*(1^+)$, $D^* \bar D^*(0^+)$, $B\bar B(0^+)$,
$B\bar B^*(1^+)$, $B^* \bar B^*(0^+)$ respectively. Apparently when
the parameters $\Lambda$ and $g_{_{PVP}}$($g_{_{VVV}}$) are
reasonable, the corresponding BS equation is solvable. For the bound
states of isospin $I=0$   we can obtain the ground eigenvalues (see
Tab. \ref{tab:decay1}).  It implies that the two heavy mesons can
form an isospin-0 molecular state by exchanging light mesons.
However for $I=1$ we cannot obtain eigenvalues which are below the
threshold with the same parameters. The reason is that the
contributions of exchanging $\rho$ and $\omega $  nearly cancel each
other in the kernel $K$.

\begin{table}
\caption{ The eigenvalues of the ground  molecular states ($I=0$)
when only $\rho$ and $\omega$ are exchanged between the
constituents} \label{tab:decay1}
\begin{tabular}{c|c|c|c|c|c|c}\hline\hline
 ~~~~~~~~~~~~~~   &  ~~~~~~$D\bar D$~~~   &
 ~~~$D\bar D^*$~~~ &  ~~~$D^*\bar D^*$~~~&  ~~~$B\bar B$~~~&  ~~~$B\bar B^*$~~~&  ~~~$B^*\bar B^*$~~~   \\
\hline
 $\Lambda=1.5$\footnote{the coupling constants $g_{_{PVP(VVV)}}$ are set to be 2}   & -    &    3.877     & 3.406 &  10.50  &  10.54  &  9.904   \\
 $\Lambda=2.5$$^a$   & 3.708    & 3.844      &  2.699  &  10.38   &   10.43  &  9.245\\
 $\Lambda=3.5$$^a$   & 3.680    & 3.815     &  2.350    & 10.33  &   10.37  &  8.950\\
 \hline
 $\Lambda=1.5$\footnote{the coupling constants $g_{_{PVP(VVV)}}$ are set to be 3}   & 3.618    &  3.756       &  3.892 &  10.26  &  10.31    &  8.899   \\
 $\Lambda=2.5$$^b$    & 3.379    &  3.529      &  3.845   &  9.955   &   10.00  &  7.771\\
 $\Lambda=3.5$$^b$  & 3.254    &  3.414      &  3.841    &  9.800   &   9.857  &  7.283\\
\hline
 $\Lambda=1.5$\footnote{the coupling constants $g_{_{PVP(VVV)}}$ are set to be 4}   & 3.330    &  3.480       &  3.637  &  9.909  &   9.956  &  7.798\\
 $\Lambda=2.5$$^c$  & 2.860    &  3.067       &  3.878  &  9.392   &   9.446  &  6.356\\
  $\Lambda=3.5$$^c$  & 2.656   &  2.886      &  3.935  &  9.162   &   9.219  &  5.802\\
 \hline\hline
 \end{tabular}
\end{table}

\subsection{As exchange of $\pi$ between two vectors is taken into account }
For the bound states composed of two vectors such as $B^*\bar B^*$
and $D^*\bar D^*$, the contribution of pion-exchange should be
included. Including  $\pi-$exchange in the kernel, we repeat our
derivation and obtain a new BS equation.  Using the data of
$D^{*+}\rightarrow D^0\pi^+$\cite{PDG10} and the formula given in
Ref. \cite{Casalbuoni:1996pg} we fix $g_{D^*\bar D\pi}=18$, and
furthermore under the heavy quark symmetry we may set $g_{B^*\bar
B^*\pi}=g_{D^*\bar D^*\pi}=g_{D^*\bar D\pi}$. With this coupling
constant we calculate the eigenvalues of  the bound states
$D^*\bar D^*$ and $B^*\bar B^*$ and  the results are listed in
Tab.\ref{tab:decay4}. Comparing the values in Tab.\ref{tab:decay4}
with that in Tab.\ref{tab:decay1} we find that the masses of the
molecular states $D^*\bar D^*$ and $B^*\bar B^*(I=0)$ slightly
decrease by less than 10\% as the pion-exchange between the two
vectors is taken into account. It shows that the pion-exchange
contributes a weak attractive effect, but it is not as important
as that from the vector exchange. Moreover, for the $I=1$ bound
state the pion-exchange with $C_I=-1$ is even less important so
there is still no $I=1$ bound state as indicated by our numerical
results.
\begin{table}
\caption{ The eigenvalues of the ground molecular states when the
contribution of $\pi$ are included with  $g_{D^*\bar
D^*\pi}=g_{B^*\bar B^*\pi}=18$ and the coupling constants
$g_{_{PVP(VVV)}}$ are set to be 3.} \label{tab:decay4}
\begin{tabular}{c|c|c|c|c}\hline\hline
 ~~~~~~~~   &  ~~~$D^*\bar D^*(I=0)$~~~   &
 ~~~$B^*\bar B^*(I=0)$~~~ &  ~~~$D^*\bar D^*$(I=1)~~~&  ~~~$B^*\bar B^*(I=1)$~~~   \\\hline
 $\Lambda=1.5$    & 3.588   & 8.489     &  -   &  -  \\
 $\Lambda=2.5$    & 3.606    &  7.270      &  -   &  -  \\
  $\Lambda=3.5$    &3.640    &  6.760      &  -   &  -  \\
\hline\hline
\end{tabular}
\end{table}

\begin{table} \caption{ The masses of the ground molecular
states when one set $\Lambda=m+\alpha \Lambda_{QCD}$ with
$g_{B\bar B\rho}=g_{B\bar B\omega}= 3$ and $g_{D^*\bar
D^*\pi}=g_{B^*\bar B^*\pi}$  }\label{tab:decay8}
\begin{tabular}{c|c|c|c|c|c|c}\hline\hline
 ~~~~~~~~   & ~~~$B\bar B$~~~&  ~~~$B\bar B^*$~~~&  ~~~$B^*\bar B^*$
 \footnote{the contribution of exchanged $\pi$ has been included.}~~~
 &  ~~~$B\bar B$~~~&  ~~~$B\bar B^*$~~~&  ~~~$B^*\bar B^*$$^a$~~~\\
 ~~~~~~   & ~~~($I=0$)~~~&  ~~~($I=0$)~~~&  ~~~($I=0$)~~~
 &  ~~~($I=1$)~~~&  ~~~($I=1$)~~~& ~~~($I=1$)~~~   \\\hline
 $\alpha=1$    & 10.49    &  10.53       &  9.939  &  -  &   -  &  -\\
 $\alpha=2$    & 10.39    &  10.44       &  9.364  &  -   &  -  &  -\\
 \hline\hline
\end{tabular}
\end{table}

As aforementioned one also can set $\Lambda=m+\alpha \Lambda_{QCD}$
which would determine values of $\Lambda$ lower than that we have
used and the relation implies different $\Lambda$ values for
different exchanged mesons. We calculate the masses of the
$B^{(*)}\bar B^{(*)}$ system for $\alpha=1$ and $\alpha=2$
respectively and the numerical results are presented in
Tab.\ref{tab:decay8}. The results in Tab.\ref{tab:decay8} are
qualitatively consistent with the conclusions we have made.

\subsection{As exchange of $\sigma$ is taken into account }
Moreover, two pseudoscalars and two vectors can couple to a scalar,
so that one should include the contribution of $\sigma-$exchange in
the kernel. In terms of the effective Lagrangian\cite{Lee:2009hy} we
deduce the corresponding kernel (See appendix). With the
contribution of $\sigma$ the eigenvalues of the bound state $D\bar
D(0^+)$, $D\bar D^*(1^+)$, $D^* \bar D^*(0^+)$, $B\bar B(0^+)$,
$B\bar B^*(1^+)$, $B^* \bar B^*(0^+)$ are re-calculated.  The
parameter $g_\sigma$ is determined by the heavy hadron chiral
perturbation theory (HHChPT) as
$g_{\sigma}=\frac{g_\pi}{2\sqrt{6}}\approx0.76$ with
$g_\pi=3.73$\footnote{We thank Dr. Valery E. Lyubovitskij  for
explaining how to determine the value of $g_\pi$.}
\cite{Lee:2009hy}. Using this value of $g_\sigma$ we repeat our
computations and the resultant masses of the ground molecular states
are listed in Tab.\ref{tab:decay5}. Comparing Tab.\ref{tab:decay1}
with Tab.\ref{tab:decay5} we  find the contribution of $\sigma$ is
very small for the concerned case.

\begin{table}
\caption{ The masses of the ground molecular states and the coupling
constants $g_{_{PVP(VVV)}}$ are set to be 3 when the contribution of
$\sigma$ are included.} \label{tab:decay5}
\begin{tabular}{c|c|c|c|c|c|c}\hline\hline
 ~~~~~~~~   &  ~~~$D\bar D$~~~   &
 ~~~$D\bar D^*$~~~ &  ~~~$D^*\bar D^*$\footnote{the contribution of
exchanging $\pi$ is not included.}~~~&  ~~~$B\bar B$~~~& ~~~$B\bar
B^*$~~~&  ~~~$B^*\bar B^*$$^a$~~~
 \\
 \hline
 $\Lambda=2.5,\,g_\sigma=0.76\,(I=0)$    & 3.386    &  3.538      &  3.839   &  9.967  &    9.990  &  7.750\\
\hline
 $\Lambda=2.5,\,g_\sigma=0.76\,(I=1)$    & -    &  -      &  -   &  -   &   -  &  -\\ \hline
 $\Lambda=2.5,\,g_\sigma=3\,(I=0)$    & 3.505    &  3.382      &  3.750   &  10.15  &    9.801  &  7.130\\
 \hline
 $\Lambda=2.5,\,g_\sigma=3\,(I=1)$    & -    &  -      &  3.760   &  -   &   10.56  &  10.22\\
 \hline
\hline
\end{tabular}
\end{table}

The above numerical results show that one cannot obtain an $I=1$
bound state which is composed of two heavy mesons. The reason is
the contributions from exchanges of $\rho$ and $\omega$ cancel
each other for the quantum number $I=1$, and the small coupling
constant $g_\sigma$ determines that including the $\sigma$
contribution cannot make a substantial change  to the kernel. Let
us take another angle to look at this issue. With the
understanding, now let us deliberately enhance the contribution of
$\sigma$ by enlarging its coupling with pseudoscalar and vector
mesons. Namely, we investigate to what value the coupling constant
$g_\sigma$ reaches, the two heavy mesons can form a molecular
state. We notice that  as the coupling constant $g_\sigma=3$ the
eigenvalues for $B\bar B^*(I=1,J^P=1^+)$, $B^*\bar
B^*(I=1,J^P=0^+)$ and $D^* \bar D^*(I=1,J^P=0^+)$  meet the
criterion for forming stable molecular states and the results are
shown in Tab.\ref{tab:decay5}. Moreover, as we set the coupling
constant $g_\sigma=4$ and $\Lambda=2.5$ we can also obtain the
mass of 3.864 GeV for $D\bar D^*(I=1,J^P=1^+)$. It is worth
emphasizing that when we set $\Lambda=2.5$ GeV and $g_\sigma=3$
the mass of the bound state $B\bar B^*(I=1,\, J^P=1^+)$ is 10.56
GeV which is close to the mass of the newly observed
narrow-structure resonance $Z_b(10610)$ [$(10608.4\pm 2.0)$ MeV].




\section{A brief summary}
In this work we study the possibility of two heavy mesons  forming a
hadronic molecule. We employ the BS framework to approach this goal
because it may include the  relativistic corrections automatically.
In Ref. \cite{Feng:2011zzb} the BS equation for the bound state of
two pseudoscalar mesons was deduced whereas in this work we derive
the BS equations of the bound states composed of one vector and one
pseudoscalar with  $J^P=1^+$ and  the bound states composed of two
vectors with  $J^P=0^+$. Our numerical results indicate that when
the parameters are within reasonable ranges two mesons indeed can
form an $I=0$ molecular state but cannot form an $I=1$ molecular
state because the contribution of $\sigma-$meson is small, thus we
conclude that the newly observed $Z_b(10610)$ prefers a tetraquark
structure or at most has a tiny fraction of the molecular state of
$B\bar B^*$. By contrast, if the coupling constant $g_\sigma$ is
enhanced by several times an $I=1$ molecular state might be formed.
For example when we set $\Lambda=2.5$ GeV and $g_\sigma=3$ the mass
of the $I=1$ bound state $B\bar B^*(1^+)$ is 10.56 GeV which is
close to the mass of the  newly observed narrow-structure resonance
$Z_b(10610)$.

Therefore our qualitative conclusion is that either the observed
$Z_b(10610)$ is not a molecular state or the coupling of
$\sigma-$meson with the pseudoscalar and vector mesons is several
times larger than that given in the HHChPT.

Since the parameters are fixed in various experiments and can span a
relatively large range we cannot expect all the numerical results
which depend on the parameters, to be very accurate. The goal of
this work is to study the possibility of two heavy mesons can form a
molecular state. Our results, even not accurate, have obvious
qualitative significance.

In a recent work, Ali et al. \cite{Ali} propose the tetraquark
interpretation of the charged bottomonium-like states
$Z_b^{\pm}(10610)$ and $Z_b^{\pm}(10650)$, and find that with this
ansatz their estimates on the production and decay fit the data
well, thus they claim that the tetraquark interpretation is
supported by the experimental measurments. This is consistent with
our conclusion that the tetraquark structure of $Z_b^{\pm}(10610)$
is preferred. Definitely, further theoretical and experimental works
are badly needed for gaining better understanding of the exotic
structures of the heavy mesons.

\section*{Acknowledgments}

We thank Dr. W. Wang for drawing our attention to their work
\cite{Ali} and we are glad to notice that the same conclusion is
reached from completely different angles. This work is supported by
the National Natural Science Foundation of China (NNSFC) under the
contract No. 11075079, No. 11005079 and No. 11175051; the Special
Grant for the Ph.D. program of Ministry of Eduction of P.R. China
No. 20100032120065.
\appendix

\section{Notations}
The effective interactions
are\cite{Feng:2011zzb,Meng:2007tk,Lee:2009hy}
\begin{eqnarray}
&&\mathcal{L}_{MM\rho}=ig_{MM\rho}[\bar M
\overrightarrow{\tau}(\partial_\mu M)-(\partial_\mu \bar M)
\overrightarrow{\tau}M ]\cdot \overrightarrow{\rho}^\mu\nonumber\\
&&\mathcal{L}_{MM\omega}=ig_{MM\omega}[\bar M (\partial_\mu
M)-(\partial_\mu \bar M) M ]\cdot \omega^\mu
\nonumber\\&&\mathcal{L}_{\rho M^*M^*}= ig_{\rho
M^*M^*}\overrightarrow{\rho}^\mu \cdot[ -
M^{*\nu}\overrightarrow{\tau}(\overrightarrow{\partial}_\mu-\overleftarrow{\partial}_\mu)
{M}_\nu^{*\dagger} +
M^{*\nu}\overrightarrow{\tau}\partial_\nu{M}^{*\dagger}_{\mu} -
\partial^\nu M_{*\mu} \overrightarrow{\tau}{M}_{*\nu\dagger}],
\nonumber\\&&\mathcal{L}_{\omega M^*M^*}= ig_{\omega M^*M^*} [
-\omega^\mu
M^{*\nu}(\overrightarrow{\partial}_\mu-\overleftarrow{\partial}_\mu)
{M}_\nu^{*\dagger} + \omega^\mu
M^{*\nu}\partial_\nu{M}^{*\dagger}_{\mu} - \omega_\mu\partial_\nu
M^{*\mu} {M}^{*\nu\dagger}],
\nonumber\\
&& \mathcal{L}_{M^*M^*\pi}=
\frac{g_{M^*M^*\pi}}{m_{M^*}}\varepsilon^{\mu\nu\alpha\beta}\partial_{\mu}M^*_{\nu}\overrightarrow{\tau}{M^*}^{\dagger}_{\alpha}
\cdot\partial_\beta\overrightarrow{\pi}
\nonumber\\
&& \mathcal{L}_{M^*M^*\sigma}=2m_{M^*}g_{\sigma} M^*\cdot
{M^*}^{\dagger}\sigma
\nonumber\\
&& \mathcal{L}_{MM\sigma}=-2m_{M}g_{\sigma} M\cdot
{M}^{\dagger}\sigma
\end{eqnarray}
and kernel

\begin{eqnarray}
K_{2V}(\mathbf{p},\mathbf{p}')=&&\frac{{ic_I g_1}\,{g_1'}}{3\,
   ( -{{M_V}}^2 + q^2 ) }\,[ 4\,{\mathbf{p}\cdot\mathbf{p'}} - \frac{{{\mathbf{p}\cdot\mathbf{p'}}}^2}{{{m_2}}^2} + 2\,{\mathbf{p'}^2} -
      \frac{{{\mathbf{p}\cdot\mathbf{p'}}}^2\,{\mathbf{p'}^2}}{{{m_2}}^2\,{{M_V}}^2} + \frac{2\,{{\mathbf{p'}^2}}^2}{{{M_V}}^2} + 2\,{\mathbf{p}^2}\nonumber\\&& +
      \frac{{{\mathbf{p}\cdot\mathbf{p'}}}^2\,{\mathbf{p}^2}}{{{m_2}}^2\,{{M_V}}^2} + \frac{{\mathbf{p'}^2}\,{\mathbf{p}^2}}{{{m_2}}^2} -
      \frac{4\,{\mathbf{p'}^2}\,{\mathbf{p}^2}}{{{M_V}}^2} + \frac{{{\mathbf{p'}^2}}^2\,{\mathbf{p}^2}}{{{m_2}}^2\,{{M_V}}^2} +
      \frac{2\,{{\mathbf{p}^2}}^2}{{{M_V}}^2} - \frac{{\mathbf{p'}^2}\,{{\mathbf{p}^2}}^2}{{{m_2}}^2\,{{M_V}}^2} +
      \frac{{{E}}^2\,{\mathbf{p}\cdot\mathbf{p'}}\,{{\eta_2}}^2}{{{m_2}}^2} \nonumber\\&& +
      \frac{{{E}}^2\,{\mathbf{p}\cdot\mathbf{p'}}\,{\mathbf{p'}^2}\,{{\eta_2}}^2}{{{m_2}}^2\,{{M_V}}^2} +
      \frac{{{E}}^2\,{\mathbf{p}^2}\,{{\eta_2}}^2}{{{m_2}}^2} -
      \frac{{{E}}^2\,{\mathbf{p}\cdot\mathbf{p'}}\,{\mathbf{p}^2}\,{{\eta_2}}^2}{{{m_2}}^2\,{{M_V}}^2} -
      \frac{{{E}}^2\,{\mathbf{p'}^2}\,{\mathbf{p}^2}\,{{\eta_2}}^2}{{{m_2}}^2\,{{M_V}}^2}\nonumber\\&&  +
      \frac{{{E}}^2\,{{\mathbf{p}^2}}^2\,{{\eta_2}}^2}{{{m_2}}^2\,{{M_V}}^2} +
      2\,{\eta_1}\,( 6\,{{E}}^2\,{\eta_2} +
      \frac{{{E}}^2\,{\mathbf{p}^2}\,{\eta_2}}{{{m_2}}^2})],
\end{eqnarray}

\begin{eqnarray}
K_{3P}(\mathbf{p},\mathbf{p}')=\frac{2c_I\,{g_2}\,{g_2'}}{{{m_P}}^2
- q^2}\,\left[ {{\mathbf{p}\cdot\mathbf{p'}}}^2 +
2\,{{E}}^2\,{\mathbf{p}\cdot\mathbf{p'}}\,{\eta 1}\,{\eta 2} -
      {{E}}^2\,{{\bf p'}}^2\,{\eta_1}\,{\eta_2} - \mathbf{p}^2\,\left( {{\bf p'}}^2 + {{E}}^2\,{\eta_1}\,{\eta_2} \right)
      \right],
\end{eqnarray}

\begin{eqnarray}
K_{3V}(\mathbf{p},\mathbf{p}')=&&\frac{{ic_I g_3}\,{g_3'}}{-{
        {M_V}}^2 + q^2}\,[ -6\,{\mathbf{p}\cdot\mathbf{p'}} - \frac{2\,{{\mathbf{p}\cdot\mathbf{p'}}}^2}{{{m_1}}^2} - \frac{2\,{{\mathbf{p}\cdot\mathbf{p'}}}^2}{{{m_2}}^2} +
      \frac{2\,{{\mathbf{p}\cdot\mathbf{p'}}}^2}{{{M_V}}^2} - 6\,{\mathbf{p'}^2} + \frac{{{\mathbf{p}\cdot\mathbf{p'}}}^2\,{\mathbf{p'}^2}}{{{m_1}}^2\,{{M_V}}^2} \nonumber\\&&+
      \frac{{{\mathbf{p}\cdot\mathbf{p'}}}^2\,{\mathbf{p'}^2}}{{{m_2}}^2\,{{M_V}}^2} - \frac{3\,{{\mathbf{p'}^2}}^2}{{{M_V}}^2} - 6\,{\mathbf{p}^2} -
      \frac{2\,{{\mathbf{p}\cdot\mathbf{p'}}}^2\,{\mathbf{p}^2}}{{{m_1}}^2\,{{m_2}}^2} - \frac{{\mathbf{p'}^2}\,{\mathbf{p}^2}}{{{m_1}}^2} -
      \frac{{\mathbf{p'}^2}\,{\mathbf{p}^2}}{{{m_2}}^2} + \frac{4\,{\mathbf{p'}^2}\,{\mathbf{p}^2}}{{{M_V}}^2} +\nonumber\\&&
      \frac{{{\mathbf{p}\cdot\mathbf{p'}}}^2\,{\mathbf{p'}^2}\,{\mathbf{p}^2}}{{{m_1}}^2\,{{m_2}}^2\,{{M_V}}^2} -
      \frac{{{\mathbf{p'}^2}}^2\,{\mathbf{p}^2}}{{{m_1}}^2\,{{M_V}}^2} - \frac{{{\mathbf{p'}^2}}^2\,{\mathbf{p}^2}}{{{m_2}}^2\,{{M_V}}^2} -
      \frac{3\,{{\mathbf{p}^2}}^2}{{{M_V}}^2} - \frac{{\mathbf{p'}^2}\,{{\mathbf{p}^2}}^2}{{{m_1}}^2\,{{m_2}}^2} -
      \frac{{{\mathbf{p'}^2}}^2\,{{\mathbf{p}^2}}^2}{{{m_1}}^2\,{{m_2}}^2\,{{M_V}}^2} +\nonumber\\&&
      \frac{2\,{{E}}^2\,{\mathbf{p}\cdot\mathbf{p'}}\,{{\eta_1}}^2}{{{m_1}}^2} +
      \frac{{{E}}^2\,{\mathbf{p'}^2}\,{{\eta_1}}^2}{{{m_1}}^2} +
      \frac{{{E}}^2\,{{\mathbf{p'}^2}}^2\,{{\eta_1}}^2}{{{m_1}}^2\,{{M_V}}^2} +
      \frac{2\,{{E}}^2\,{\mathbf{p}^2}\,{{\eta_1}}^2}{{{m_1}}^2} -
      \frac{2\,{{E}}^2\,{\mathbf{p'}^2}\,{\mathbf{p}^2}\,{{\eta_1}}^2}{{{m_1}}^2\,{{M_V}}^2} \nonumber\\&&+
      \frac{{{E}}^2\,{{\mathbf{p}^2}}^2\,{{\eta_1}}^2}{{{m_1}}^2\,{{M_V}}^2} -
      \frac{{{E}}^2\,{\mathbf{p}\cdot\mathbf{p'}}\,{\eta_1}\,{\eta_2}}{{{m_1}}^2} -
      \frac{2\,{{E}}^2\,{{\mathbf{p}\cdot\mathbf{p'}}}^2\,{\eta_1}\,{\eta_2}}{{{m_1}}^2\,{{m_2}}^2}  +
      \frac{{{E}}^2\,{{\mathbf{p}\cdot\mathbf{p'}}}^2\,{\eta_1}\,{\eta_2}}{{{m_1}}^2\,{{M_V}}^2} \nonumber\\&&+
      \frac{{{E}}^2\,{\mathbf{p}\cdot\mathbf{p'}}\,{\mathbf{p'}^2}\,{\eta_1}\,{\eta_2}}{{{m_1}}^2\,{{M_V}}^2} +
      \frac{{{E}}^2\,{{\mathbf{p}\cdot\mathbf{p'}}}^2\,{\mathbf{p'}^2}\,{\eta_1}\,{\eta_2}}{{{m_1}}^2\,{{m_2}}^2\,{{M_V}}^2}+
      \frac{3\,{{E}}^2\,{\mathbf{p}^2}\,{\eta_1}\,{\eta_2}}{{{m_1}}^2} -
      \frac{{{E}}^2\,{\mathbf{p}\cdot\mathbf{p'}}\,{\mathbf{p}^2}\,{\eta_1}\,{\eta_2}}{{{m_1}}^2\,{{m_2}}^2}\nonumber\\&& -
      \frac{{{E}}^2\,{\mathbf{p}\cdot\mathbf{p'}}\,{\mathbf{p}^2}\,{\eta_1}\,{\eta_2}}{{{m_1}}^2\,{{M_V}}^2} -
      \frac{2\,{{E}}^2\,{\mathbf{p'}^2}\,{\mathbf{p}^2}\,{\eta_1}\,{\eta_2}}{{{m_1}}^2\,{{m_2}}^2} -
      \frac{2\,{{E}}^2\,{\mathbf{p'}^2}\,{\mathbf{p}^2}\,{\eta_1}\,{\eta_2}}{{{m_1}}^2\,{{M_V}}^2} +
      \frac{{{E}}^2\,{\mathbf{p}\cdot\mathbf{p'}}\,{\mathbf{p'}^2}\,{\mathbf{p}^2}\,{\eta_1}\,{\eta_2}}{{{m_1}}^2\,{{m_2}}^2\,{{M_V}}^2} \nonumber\\&& -
      \frac{2\,{{E}}^2\,{{\mathbf{p'}^2}}^2\,{\mathbf{p}^2}\,{\eta_1}\,{\eta_2}}{{{m_1}}^2\,{{m_2}}^2\,{{M_V}}^2} +
      \frac{{{E}}^2\,{{\mathbf{p}^2}}^2\,{\eta_1}\,{\eta_2}}{{{m_1}}^2\,{{m_2}}^2} +
      \frac{{{E}}^2\,{{\mathbf{p}^2}}^2\,{\eta_1}\,{\eta_2}}{{{m_1}}^2\,{{M_V}}^2} -
      \frac{{{E}}^2\,{\mathbf{p}\cdot\mathbf{p'}}\,{{\mathbf{p}^2}}^2\,{\eta_1}\,{\eta_2}}{{{m_1}}^2\,{{m_2}}^2\,{{M_V}}^2} \nonumber\\&&+
      \frac{{{E}}^2\,{\mathbf{p'}^2}\,{{\mathbf{p}^2}}^2\,{\eta_1}\,{\eta_2}}{{{m_1}}^2\,{{m_2}}^2\,{{M_V}}^2} +
      \frac{3\,{{E}}^2\,{\mathbf{p}\cdot\mathbf{p'}}\,{{\eta_2}}^2}{{{m_2}}^2} -
      \frac{{{E}}^2\,{{\mathbf{p}\cdot\mathbf{p'}}}^2\,{{\eta_2}}^2}{{{m_2}}^2\,{{M_V}}^2}+
      \frac{{{E}}^2\,{\mathbf{p'}^2}\,{{\eta_2}}^2}{{{m_2}}^2} -
      \frac{{{E}}^2\,{\mathbf{p}\cdot\mathbf{p'}}\,{\mathbf{p'}^2}\,{{\eta_2}}^2}{{{m_2}}^2\,{{M_V}}^2}\nonumber\\&& +
      \frac{{{E}}^2\,{{\mathbf{p'}^2}}^2\,{{\eta_2}}^2}{{{m_2}}^2\,{{M_V}}^2}  -
      \frac{{{E}}^2\,{\mathbf{p}^2}\,{{\eta_2}}^2}{{{m_2}}^2} +
      \frac{2\,{{E}}^2\,{\mathbf{p}\cdot\mathbf{p'}}\,{\mathbf{p}^2}\,{{\eta_2}}^2}{{{m_1}}^2\,{{m_2}}^2} +
      \frac{{{E}}^2\,{\mathbf{p}\cdot\mathbf{p'}}\,{\mathbf{p}^2}\,{{\eta_2}}^2}{{{m_2}}^2\,{{M_V}}^2} -
      \frac{{{E}}^2\,{{\mathbf{p}\cdot\mathbf{p'}}}^2\,{\mathbf{p}^2}\,{{\eta_2}}^2}{{{m_1}}^2\,{{m_2}}^2\,{{M_V}}^2}\nonumber\\&&+
      \frac{{{E}}^2\,{\mathbf{p'}^2}\,{{\mathbf{p}^2}}^2\,{{\eta_2}}^2}{{{m_1}}^2\,{{m_2}}^2\,{{M_V}}^2} -
      \frac{{{E}}^4\,{\mathbf{p}\cdot\mathbf{p'}}\,{{\eta_1}}^2\,{{\eta_2}}^2}{{{m_1}}^2\,{{m_2}}^2} -
      \frac{{{E}}^4\,{\mathbf{p'}^2}\,{{\eta_1}}^2\,{{\eta_2}}^2}{{{m_1}}^2\,{{m_2}}^2} +
      \frac{{{E}}^4\,{\mathbf{p}\cdot\mathbf{p'}}\,{\mathbf{p'}^2}\,{{\eta_1}}^2\,{{\eta_2}}^2}{{{m_1}}^2\,{{m_2}}^2\,{{M_V}}^2} \nonumber\\&& -
      \frac{{{E}}^4\,{{\mathbf{p'}^2}}^2\,{{\eta_1}}^2\,{{\eta_2}}^2}{{{m_1}}^2\,{{m_2}}^2\,{{M_V}}^2}+
      \frac{{{E}}^4\,{\mathbf{p}^2}\,{{\eta_1}}^2\,{{\eta_2}}^2}{{{m_1}}^2\,{{m_2}}^2} -
      \frac{{{E}}^4\,{\mathbf{p}\cdot\mathbf{p'}}\,{\mathbf{p}^2}\,{{\eta_1}}^2\,{{\eta_2}}^2}{{{m_1}}^2\,{{m_2}}^2\,{{M_V}}^2} +
      \frac{{{E}}^4\,{\mathbf{p'}^2}\,{\mathbf{p}^2}\,{{\eta_1}}^2\,{{\eta_2}}^2}{{{m_1}}^2\,{{m_2}}^2\,{{M_V}}^2} \nonumber\\&&+
      \frac{2\,{{E}}^4\,{\mathbf{p}\cdot\mathbf{p'}}\,{\eta_1}\,{{\eta_2}}^3}{{{m_1}}^2\,{{m_2}}^2} -
      \frac{{{E}}^4\,{{\mathbf{p}\cdot\mathbf{p'}}}^2\,{\eta_1}\,{{\eta_2}}^3}{{{m_1}}^2\,{{m_2}}^2\,{{M_V}}^2} +
      \frac{{{E}}^4\,{\mathbf{p'}^2}\,{\mathbf{p}^2}\,{\eta_1}\,{{\eta_2}}^3}{{{m_1}}^2\,{{m_2}}^2\,{{M_V}}^2} -
      2\,{\eta_1}\,( \frac{{{E}}^2\,{\mathbf{p}\cdot\mathbf{p'}}\,{\eta_1}}{{{m_1}}^2}\nonumber\\&& +
         \frac{{{E}}^2\,{\mathbf{p}^2}\,{\eta_1}}{{{m_1}}^2} +
         \frac{{{E}}^2\,{\mathbf{p}\cdot\mathbf{p'}}\,{\mathbf{p}^2}\,{\eta_1}}{{{m_1}}^2\,{{m_2}}^2} + 6\,{{E}}^2\,{\eta_2} -
         \frac{{{E}}^2\,{\mathbf{p}\cdot\mathbf{p'}}\,{\eta_2}}{{{m_2}}^2} + \frac{2\,{{E}}^2\,{\mathbf{p}^2}\,{\eta_2}}{{{m_1}}^2} +
         \frac{{{E}}^2\,{\mathbf{p}^2}\,{\eta_2}}{{{m_2}}^2} \nonumber\\&& +
         \frac{{{E}}^2\,{{\mathbf{p}^2}}^2\,{\eta_2}}{{{m_1}}^2\,{{m_2}}^2}+
         \frac{{{E}}^4\,{\mathbf{p}\cdot\mathbf{p'}}\,{{\eta_1}}^2\,{\eta_2}}{{{m_1}}^2\,{{m_2}}^2} +
         \frac{{{E}}^4\,{\mathbf{p}^2}\,{\eta_1}\,{{\eta_2}}^2}{{{m_1}}^2\,{{m_2}}^2} )  +
      {\eta_2}\,[ - \frac{{{E}}^2\,{\mathbf{p}\cdot\mathbf{p'}}\,{\mathbf{p'}^2}\,{\eta_1}}{{{m_1}}^2\,{{M_V}}^2}   -
         \frac{{{E}}^2\,{\mathbf{p}^2}\,{\eta_1}}{{{m_1}}^2}\nonumber\\&& +
         \frac{{{E}}^2\,{\mathbf{p}\cdot\mathbf{p'}}\,{\mathbf{p}^2}\,{\eta_1}}{{{m_1}}^2\,{{m_2}}^2} +
         \frac{{{E}}^2\,{\mathbf{p}\cdot\mathbf{p'}}\,{\mathbf{p}^2}\,{\eta_1}}{{{m_1}}^2\,{{M_V}}^2}+
         \frac{{{E}}^2\,{\mathbf{p'}^2}\,{\mathbf{p}^2}\,{\eta_1}}{{{m_1}}^2\,{{M_V}}^2} -
         \frac{{{E}}^2\,{{\mathbf{p}^2}}^2\,{\eta_1}}{{{m_1}}^2\,{{M_V}}^2} - 3\,{{E}}^2\,{\eta_2} \nonumber\\&&+
         \frac{2\,{{E}}^2\,{\mathbf{p}\cdot\mathbf{p'}}\,{\eta_2}}{{{M_V}}^2} -
         \frac{{{E}}^2\,{{\mathbf{p}\cdot\mathbf{p'}}}^2\,{\eta_2}}{{{m_1}}^2\,{{M_V}}^2} -
         \frac{{{E}}^2\,{{\mathbf{p}\cdot\mathbf{p'}}}^2\,{\eta_2}}{{{m_2}}^2\,{{M_V}}^2} -
         \frac{{{E}}^2\,{\mathbf{p'}^2}\,{\eta_2}}{{{M_V}}^2}+
         \frac{{{E}}^2\,{\mathbf{p}\cdot\mathbf{p'}}\,{\mathbf{p'}^2}\,{\eta_2}}{{{m_2}}^2\,{{M_V}}^2} \nonumber\\&&-
         \frac{{{E}}^2\,{\mathbf{p}^2}\,{\eta_2}}{{{m_1}}^2} - \frac{{{E}}^2\,{\mathbf{p}^2}\,{\eta_2}}{{{M_V}}^2} +
         \frac{{{E}}^2\,{\mathbf{p}\cdot\mathbf{p'}}\,{\mathbf{p}^2}\,{\eta_2}}{{{m_1}}^2\,{{m_2}}^2} +
         \frac{2\,{{E}}^2\,{\mathbf{p}\cdot\mathbf{p'}}\,{\mathbf{p}^2}\,{\eta_2}}{{{m_1}}^2\,{{M_V}}^2} +
         \frac{{{E}}^2\,{\mathbf{p}\cdot\mathbf{p'}}\,{\mathbf{p}^2}\,{\eta_2}}{{{m_2}}^2\,{{M_V}}^2} -\nonumber\\&&
         \frac{{{E}}^2\,{{\mathbf{p}\cdot\mathbf{p'}}}^2\,{\mathbf{p}^2}\,{\eta_2}}{{{m_1}}^2\,{{m_2}}^2\,{{M_V}}^2} -
         \frac{{{E}}^2\,{\mathbf{p'}^2}\,{\mathbf{p}^2}\,{\eta_2}}{{{m_2}}^2\,{{M_V}}^2} +
         \frac{{{E}}^2\,{\mathbf{p}\cdot\mathbf{p'}}\,{\mathbf{p'}^2}\,{\mathbf{p}^2}\,{\eta_2}}{{{m_1}}^2\,{{m_2}}^2\,{{M_V}}^2}  -
         \frac{{{E}}^2\,{{\mathbf{p}^2}}^2\,{\eta_2}}{{{m_1}}^2\,{{M_V}}^2}+
         \frac{{{E}}^2\,{\mathbf{p}\cdot\mathbf{p'}}\,{{\mathbf{p}^2}}^2\,{\eta_2}}{{{m_1}}^2\,{{m_2}}^2\,{{M_V}}^2}\nonumber\\&& -
         \frac{{{E}}^2\,{\mathbf{p'}^2}\,{{\mathbf{p}^2}}^2\,{\eta_2}}{{{m_1}}^2\,{{m_2}}^2\,{{M_V}}^2} +
         \frac{{{E}}^4\,{\mathbf{p}\cdot\mathbf{p'}}\,{{\eta_1}}^2\,{\eta_2}}{{{m_1}}^2\,{{m_2}}^2} +
         \frac{{{E}}^4\,{\mathbf{p}\cdot\mathbf{p'}}\,{\eta_1}\,{{\eta_2}}^2}{{{m_1}}^2\,{{m_2}}^2} -
         \frac{{{E}}^4\,{{\mathbf{p}\cdot\mathbf{p'}}}^2\,{\eta_1}\,{{\eta_2}}^2}{{{m_1}}^2\,{{m_2}}^2\,{{M_V}}^2} \nonumber\\&&+
         \frac{{{E}}^4\,{\mathbf{p}\cdot\mathbf{p'}}\,{\mathbf{p'}^2}\,{\eta_1}\,{{\eta_2}}^2}{{{m_1}}^2\,{{m_2}}^2\,{{M_V}}^2} +
         \frac{{{E}}^4\,{\mathbf{p}\cdot\mathbf{p'}}\,{\mathbf{p}^2}\,{\eta_1}\,{{\eta_2}}^2}{{{m_1}}^2\,{{m_2}}^2\,{{M_V}}^2} -
         \frac{{{E}}^4\,{\mathbf{p'}^2}\,{\mathbf{p}^2}\,{\eta_1}\,{{\eta_2}}^2}{{{m_1}}^2\,{{m_2}}^2\,{{M_V}}^2} ]  +
      {\eta_1}\,( \frac{3\,{{E}}^2\,{\mathbf{p}\cdot\mathbf{p'}}\,{\eta_1}}{{{m_1}}^2} \nonumber\\&&-
         \frac{{{E}}^2\,{{\mathbf{p}\cdot\mathbf{p'}}}^2\,{\eta_1}}{{{m_1}}^2\,{{M_V}}^2} +
         \frac{3\,{{E}}^2\,{\mathbf{p}\cdot\mathbf{p'}}\,{\mathbf{p}^2}\,{\eta_1}}{{{m_1}}^2\,{{m_2}}^2} -
         \frac{{{E}}^2\,{{\mathbf{p}\cdot\mathbf{p'}}}^2\,{\mathbf{p}^2}\,{\eta_1}}{{{m_1}}^2\,{{m_2}}^2\,{{M_V}}^2} +
         \frac{{{E}}^2\,{\mathbf{p'}^2}\,{\mathbf{p}^2}\,{\eta_1}}{{{m_1}}^2\,{{M_V}}^2}  \nonumber\\&&+
         \frac{{{E}}^2\,{\mathbf{p'}^2}\,{{\mathbf{p}^2}}^2\,{\eta_1}}{{{m_1}}^2\,{{m_2}}^2\,{{M_V}}^2} -
         3\,{{E}}^2\,{\eta_2} - \frac{3\,{{E}}^2\,{\mathbf{p}\cdot\mathbf{p'}}\,{\eta_2}}{{{m_2}}^2}+
         \frac{2\,{{E}}^2\,{\mathbf{p}\cdot\mathbf{p'}}\,{\eta_2}}{{{M_V}}^2} +
         \frac{{{E}}^2\,{{\mathbf{p}\cdot\mathbf{p'}}}^2\,{\eta_2}}{{{m_2}}^2\,{{M_V}}^2} \nonumber\\&& -
         \frac{{{E}}^2\,{\mathbf{p'}^2}\,{\eta_2}}{{{M_V}}^2} - \frac{{{E}}^2\,{\mathbf{p}^2}\,{\eta_2}}{{{M_V}}^2}-
         \frac{{{E}}^2\,{\mathbf{p'}^2}\,{\mathbf{p}^2}\,{\eta_2}}{{{m_2}}^2\,{{M_V}}^2} +
         \frac{3\,{{E}}^4\,{{\eta_1}}^2\,{\eta_2}}{{{m_1}}^2} +
         \frac{3\,{{E}}^4\,{\mathbf{p}\cdot\mathbf{p'}}\,{{\eta_1}}^2\,{\eta_2}}{{{m_1}}^2\,{{m_2}}^2} \nonumber\\&& -
         \frac{2\,{{E}}^4\,{\mathbf{p}\cdot\mathbf{p'}}\,{{\eta_1}}^2\,{\eta_2}}{{{m_1}}^2\,{{M_V}}^2} -
         \frac{{{E}}^4\,{{\mathbf{p}\cdot\mathbf{p'}}}^2\,{{\eta_1}}^2\,{\eta_2}}{{{m_1}}^2\,{{m_2}}^2\,{{M_V}}^2}+
         \frac{{{E}}^4\,{\mathbf{p'}^2}\,{{\eta_1}}^2\,{\eta_2}}{{{m_1}}^2\,{{M_V}}^2} +
         \frac{{{E}}^4\,{\mathbf{p}^2}\,{{\eta_1}}^2\,{\eta_2}}{{{m_1}}^2\,{{M_V}}^2} \nonumber\\&& +
         \frac{{{E}}^4\,{\mathbf{p'}^2}\,{\mathbf{p}^2}\,{{\eta_1}}^2\,{\eta_2}}{{{m_1}}^2\,{{m_2}}^2\,{{M_V}}^2} -
         \frac{3\,{{E}}^4\,{\mathbf{p}^2}\,{\eta_1}\,{{\eta_2}}^2}{{{m_1}}^2\,{{m_2}}^2}  +
         \frac{2\,{{E}}^4\,{\mathbf{p}\cdot\mathbf{p'}}\,{\mathbf{p}^2}\,{\eta_1}\,{{\eta_2}}^2}{{{m_1}}^2\,{{m_2}}^2\,{{M_V}}^2}-
         \frac{{{E}}^4\,{\mathbf{p'}^2}\,{\mathbf{p}^2}\,{\eta_1}\,{{\eta_2}}^2}{{{m_1}}^2\,{{m_2}}^2\,{{M_V}}^2} \nonumber\\&&-
         \frac{{{E}}^4\,{{\mathbf{p}^2}}^2\,{\eta_1}\,{{\eta_2}}^2}{{{m_1}}^2\,{{m_2}}^2\,{{M_V}}^2} +
         \frac{3\,{{E}}^4\,{{\eta_2}}^3}{{{m_2}}^2} -
         \frac{2\,{{E}}^4\,{\mathbf{p}\cdot\mathbf{p'}}\,{{\eta_2}}^3}{{{m_2}}^2\,{{M_V}}^2} +
         \frac{{{E}}^4\,{\mathbf{p'}^2}\,{{\eta_2}}^3}{{{m_2}}^2\,{{M_V}}^2} +
         \frac{{{E}}^4\,{\mathbf{p}^2}\,{{\eta_2}}^3}{{{m_2}}^2\,{{M_V}}^2} \nonumber\\&&-
         \frac{3\,{{E}}^6\,{{\eta_1}}^2\,{{\eta_2}}^3}{{{m_1}}^2\,{{m_2}}^2} +
         \frac{2\,{{E}}^6\,{\mathbf{p}\cdot\mathbf{p'}}\,{{\eta_1}}^2\,{{\eta_2}}^3}{{{m_1}}^2\,{{m_2}}^2\,{{M_V}}^2}  -
         \frac{{{E}}^6\,{\mathbf{p'}^2}\,{{\eta_1}}^2\,{{\eta_2}}^3}{{{m_1}}^2\,{{m_2}}^2\,{{M_V}}^2} -
         \frac{{{E}}^6\,{\mathbf{p}^2}\,{{\eta_1}}^2\,{{\eta_2}}^3}{{{m_1}}^2\,{{m_2}}^2\,{{M_V}}^2} )
         ],
\end{eqnarray}

\begin{eqnarray}K_{1S}(\mathbf{p},\mathbf{p}')= \frac{4\,c_I{g_1}\,{g_1'2}\,{m_1 m_2}}
  {\left( {{m_S}}^2 - {\mathbf{q}^2} \right) },
\end{eqnarray}

\begin{eqnarray}K_{2S}(\mathbf{p},\mathbf{p}')= -\frac{4\,c_I{g_2}\,{g_2'}\,{m_1}\,\left( 3\,{{m_2}}^2 +
{\mathbf{p}^2} \right) }
  {3\,{m_2}\,\left( {{m_S}}^2 - {\mathbf{q}^2} \right) },
\end{eqnarray}

\begin{eqnarray}K_{3S}(\mathbf{p},\mathbf{p}')=\frac{-4\,c_I{g_3}\,{g_3'}\,\left[ {{m_2}}^2\,\left( {\mathbf{p}^2} -
{{ee}}^2\,{{\eta_1}}^2 \right)  +
      {\left( {\mathbf{p}^2} + {{ee}}^2\,{\eta_1}\,{\eta_2} \right) }^2 +
      {{m_1}}^2\,\left( 4\,{{m_2}}^2 + \mathbf{p}^2 - {{ee}}^2\,{{\eta_2}}^2 \right)  \right] }{{m_1}\,
    {m_2}\,\left( {{m_S}}^2 - \mathbf{q}^2 \right) },
\end{eqnarray}

where $g_1$, $g_1'$, $g_2$, $g_2'$, $g_3$ and $g_3'$ represent the
coupling constants in the vertexes  and  $\mathbf{q}^2=
(\mathbf{p}^2 + \mathbf{p}'^2 - 2 \mathbf{p}\cdot \mathbf{p}')$.

\end{document}